# Ethical Considerations of AR Applications in Smartphones; A Systematic Literature Review of Consumer Perspectives

**Author**

Nicola J Wood [1 Curtin University]

1 – MS Marketing, Grad Cert Business, Bachelor of Creative Technology

# Table of Contents

# Abstract

This study focuses on the ethical considerations that a consumer perceives with augmented reality (AR) in the context of smartphone applications. Through a systematic review, this research can provide an understanding and ability for developers, product managers, digital marketers and associated business professionals to effectively implement and deploy mobile AR related applications and campaigns, with consideration to the perceptions of the ethical considerations that consumers have of this growing technology. The rise in digital transformation and new technologies paved this research agenda.

Trends in the data revealed two overarching factors of 'Benefits' and 'Ethical Considerations'. Within these two factors, several consumer perceived themes were identified with regards to AR applications and their association categorised either positive, negative or neutral. 'Benefits' revealed 3 consistent themes of personalisation, interactivity and information acquisition. 'Ethical Considerations' revealed consistent patterns of educational awareness, privacy, transparency and security.

From identifying the consumer perceptions, business professionals can strategically address and or challenge the inherent limitations and their associations during AR application development, product adoption strategies or marketing purposes.

# 1. Introduction

## 1.1 Consumer Perceptions Towards Businesses

Consumer perceptions are a dynamic construct. The ability for businesses to build a trusting relationship with a consumer has been increasing in necessity with the advancement of digital technologies. The basic concept of marketing goods and services to society has been a common practice for many centuries. However modern marketing as we may know it began around the industrial revolution. It has gone through many iterations since then from traditional to human centred marketing. According to Bhayani (2018), consumers are becoming smarter in regard to how they purchase and desire experiences of goods or services. This change, in tandem with the advances of the internet and the plethora of information now available at our fingertips, has paved the way for digital marketing and the way consumers perceive organisations. The perceptions of consumers have changed the behaviour around purchase decisions or how they choose to interact with a business.

Organisations are having to adapt their processes in ways that enable trust and a relationship between with consumers. Terms such as customer satisfaction are now highly valuable and a measured part of a business model (Corbitt, Thanasankit, and Yi, 2003). Consumers of this growing digitally driven world are beginning to engage in a business relationship with an organisation, at times purely based off the beliefs, values and attitudes of a brand and if that resonates with them as an individual. Consumer perceptions can come from many forms such as the type of promotional activity, packaging or visual appeal, service quality and the quality of digital platforms, messaging, political stance and affiliations, sustainability efforts, corporate social responsibility, transparency on data acquisitions, security and privacy of data, social-causes, accessibility, product development and user experience. With organisations embarking on new digital transformation models, the stakes are higher towards how a consumer perceives every activity from a business as compared to conventional traditional business models (Khosla, Damiani, and Grosky 2003). This in turn forces businesses to look more explicitly at consumer perceptions.

## 1.2 Augmented Reality in Marketing

Augmented Reality (AR) is a digitally developed technology that aims to facilitate an interactive experience to the user through computer generated perceptions within the real-world (Rauschnabel et al, 2022). It is a form of the umbrella term *Immersive Technologies* of which aims to extend reality. This form of technology can be aimed at an individual or group experience and are seen in such places as large as art installations to mobile phone applications. Augmented reality largely entails video sensing components involving images, sounds and text. It has been used across many industries and for different purposes such as training and education, repair and maintenance, retail, design modelling and builds, operation and business logistics, field service and entertainment (Carmigniani et al. 2010).

The subdiscipline of 'Augmented Reality Marketing' is still a novel concept, however the adoption of this technology is increasing in business organisations. As such, AR has developed in the recent years with a strong focus on mobile applications as stated by Davidavičienė, Raudeliūnienė & Viršilaitė (2021). Augmented reality was first developed by Ivan Sutherland, a Harvard computer scientist in 1968 (Javornik, 2016). The first commercial use of AR was a paper printed magazine ad for the automobile company BMW. Appearing in 2008, the ad developed by German agencies allowed a user to position the paper over a computer camera which displayed a digital model of the car in real time (Javornik, 2016). The user could virtually interact through manoeuvring the piece of paper. This was the first marketing campaign displaying augmented reality to a consumer. Since 2008, paper-based marketing campaigns have moved to other devices such a smartphone where there has been an increase in discussion on the topic of digital ethics surrounding the applications of AR.

### 1.3 Smartphone History and Current Users

Smartphones differ from mobile phones in that both can be used for calls and text. However where the term smartphone came into effect, was based on the advances in mobile computing. The first ever handheld mobile phone and call made was in 1973 by Motorola. Weighing in at 2kg with a talk time of only 30 minutes and charge tome of 10 hours, this was an impressive feat by inventor Dr Martin Cooper (Kaur et al. 2021). What society knows as a smartphone was actually invented by IMB in 1992. The phone, called IBM Simon, involved an ability to send emails, annotate through a touch screen, a keyboard type input system, calendar with appointment scheduling, an address book as well as the ability to make calls and send texts. These features are now standard in today's smartphone that were present in IBM Simon (Islam and Want 2014). Most notably, the introduction of the Apple iPhone in 2007 revolutionised smartphone technology as we know it through their operating system iOS. This device brought with it a full internet experience similar to a desktop computer or laptop. However in late 2007, Google introduced its operating system Android with the intention to penetrate the smartphone market (Vohra 2016). Their open-source technology enabled many competitors to enter the market and build devices. Presently, devices by Apple and Samsung are the two most sought-after smartphone developers according to Statista (2022).

With the rapid growth of digital technologies in mobile computing and the internet, there are currently more than 7 billion mobile users in the world (Statista 2019). This is different to smartphone usage of which is approximately two thirds of the current world population (Statista 2020). To put the current usage into perspective, a survey was completed in the US across 21 different consumer demographics such as ethnicity, age, level of education and income. Every demographic revealed a significantly higher percentage of smartphone usage than that of a mobile phone (Silver 2019). As such, the level of consumer demand, expectations and perceptions have changed as organisations are forced to operate within the digital realm.

### 1.4 Ethics in the Digital Realm

With an ever-changing construct of the parameters in which consumers and business organisations operate, the phenomenon of digital ethics has arisen most prominently over the last decade. The practice of ethics has fundamentally been assumed to emerge when human societies developed an awareness and reflection on the best way to live (Singer 2018). This process began the first introduction of moral codes. Digital ethics however can be described as ethical dimensions among the shifting digital ecologies (Reyman and Sparby 2021). These rules and principals particularly focus on internet data and the behavioural governance to the inception and curation of this data. Digital ethics can be the study either demonstrated from an individual or an organisational level. This can be from how one conducts themselves online across various digital mediums and contexts to the larger construct of an enterprise's business model. Digital ethics for business organisations in particular are becoming just as valuable as a product or service itself where the perceptions of the conduct and moral compass of a business can influence consumers. This ethical governance can also be incorporated into the development of the products themselves. Digital ethics are no longer becoming a marketing technique or buzzword, as more and more consumers are avoiding doing business with organisations that do not protect their data (Whiting and Pritchard 2017). Some of the main digital ethical principles for digital transformation of organisations include:

- Designing products and services for integrity, privacy and security.
- Promoting trust among consumers.
- Being transparent and conscious of bias.
- Maintaining corporate accountability.
- Developing and promoting a culture of ethics.

# 2. Background - A Literature Review

The purpose of this chapter is an evaluation of the research to the background of this study. The literature review is broken down into the four main sections of this research including; *Digital Transformation in Business, Innovation of AR on Smartphones, Smartphones and Consumers* and *Ethical Concerns with Digital Technologies*. The four elements encompass the background to this overall study and include peer reviewed articles which form the basis of this literature investigation. There is an interlinked nature with the four factors of this background however have been broken down into sections that include both the positive and negative context of each.

## 2.1 Digital Transformation in Business

According to Reis et al. (2018), digital transformation is the term used to explain the technologies that aim to influence various social and economic factors in a positive manner. Digital transformation is not a new term however has been gaining increased attention in recent years. For more than 30 years we have been implementing computerised technologies in which individuals and businesses have developed a deeply embedded connection to the digital world. However, with the rapid expansion of growing technologies this has forced many organisations to evaluate their current business models. This was further propelled by events such as the Covid 19 pandemic where many business and social systems were redirected, and in parts wholly, to the digital realm (Priyono, Moin, and Putri 2020).

In this modern era, the digital transformation boom has completely re-structured some company's business models. Nearly everything one can think of, has formed a reliance with a digital element. As many businesses operate within this digital ecosystem, the need to remain an active participant in the digital realm is of great importance to business longevity and presence (Iansiti and Lakhani 2020).

One such study by Gil-Gomez et al. (2020) investigated sustainable business models through digital transformation. They revealed a single implementation of a Customer Relationship Management (CRM) system had positive effects across all areas of a business. Some of these were information sharing, customer involvement, long-term partnership and joint problem-solving through a digital medium. This one solution was suggested as a fundamental innovation to enable sustainable economic and financial growth. It enables companies to positively exploit their resources.

This positive effect towards a business is further validated by an empirical study conducted by Nwankpa and Roumani in 2016. They evaluated various chief information officers (CIO) across United States-based firms with a research model through data collected by surveys. This study was to evaluate IT capabilities that contribute to digital transformation efforts. As businesses are surrounded by many emerging digital technologies, evaluation of the firm's digital footprint on business performance is critical to understanding business growth and success. Their study found a positive correlation of digital transformation on a firm's performance as well as a positive relationship between IT capabilities and digital transformation. They emphasised the importance of having suitable business strategies in place for any digital transformation efforts when fostering a firm's performance.

However, this was not so much the debate of Calvino and Spiezia (2020). They state that although digital transformation may have the potential to improve productivity and enhance incomes, the opposite is also true. They reviewed recent studies on employment levels among information and communication technologies (ICTs) providing new evidence on labour demands and job destruction. They also found negative influences of digital transformation in relation to discrimination of sexual orientation, gender and income inequality.

Franke and Zoubir (2020) argue the importance of incorporating humanity in digital technologies. They have the perspective of which humanity ultimately developed artefacts as a means to enable resources such as food, time, health and mobility for the benefit both psychologically and physiologically for humans. They believe this perspective of humanity must continue to form the orientation in which to build further technological innovations upon. They coin this development as a truly human-centred digital transformation. Through this brief examination of the literature on digital transformation, there are various viewpoints of how this development has had a multitude of effects towards people, businesses and society in general.

### 2.2 Innovation of AR on Smartphones

Augmented reality is becoming a hot topic in today's society both for businesses and consumers. Although in the early stages of development and implementation, to some this technology is set to be a pervasive digital norm for our future. As augmented reality is the altering of the real world through immersing users in a reproduced digital reality, the application on smartphones provides a direct avenue to a user (Valentin et al. 2019). For the most part, augmented reality is a type of immerse technology where Edwards-Stewart, Tim, and Reger (2016) define two types of augmented reality based on their application. They are Triggered AR technologies and View-based AR. With triggered AR, the augmentation is as the name suggests, triggered, from a marker, location or dynamic response to an object. Triggered AR can also be complex augmentation, which is a combination of the above with information also accessed from the internet. One such example is Google Glass that includes GPS location marketers as well as information displayed about the object-based stimulus which is sourced from the internet.

View-based AR includes either an indirect or non-specific type of augmentation. This type of augmentation can alter images from a static viewpoint. An example is the Wall Painter app in which intelligently augments an image of the real world. For non-specific augmented reality the view can be dynamic and is not based on reference to the user's environment. This is a common form of AR used in mobile app games (Edwards-Stewart, Tim, and Reger 2016).

With the technology involved in developing augmented reality applications, smartphones provide several of the key factors to enabling this technology. Although augmented reality can be utilised from other digital devices such as tablets or smart glasses, smartphones are the most widely used device currently enabling these elements such as camera sensing, location tracking, access to the internet while the computer vision technology required is often updated with new device releases. As such, it provides a convenient means of the direct user interaction and experience through these elements.

With the convenient application of augmented reality on smartphones, there have been many studies conducted to evaluate human behaviour and gain insights. Sinha and Srivastava (2021) conducted a study with 458 respondents to assess the capabilities of AR on customer brand engagement (CBE). They also evaluated brand satisfaction and brand loyalty in relation to the AR application. Their studies confirmed that gamification involving AR had positive relationship with all elements of customer brand engagement. Vividness, novelty and interactivity were explored with the game elements where this also influenced the positive relationships with CBE. An example of this type of AR application is the Instagram game face filters.

Rauschnabel (2021) further extends this notion of smartphone AR applications with consumer acceptance and experience. They found, based on a study of 2000 respondents, that consumers had higher acceptance rates for holographic products using AR substitutes than for some real products. These products included things such as manuals, post-it notes or navigation technology. Consumers were less likely to accept virtual substitutes for other products such as memorabilia and pets. Certain characteristics of consumers were taken into consideration, however the study concluded with implications of a "copying and pasting" effect on real-world elements giving rise to virtual counterfeits.

Evaluating the positive and negative consequences of AR apps was conducted in a study by Smink et al. (2020). They tested two apps, an AR make-up application of the face, and an AR furniture application in ones surrounding space. They concluded that with the facial app, there was a higher negative persuasive consequence due to perceived intrusiveness, however this did enhance purchase intentions due to the personalisation. The furniture app revealed no negative persuasive consequence. This study can possibly reflect an understanding that consumers do enjoy AR apps in relation to a personal context, however there are limitations to how far this extends before negative persuasive consequences are activated.

## 2.3 Consumer Use of Smartphone Technology

As the level of smartphone usage grows to phenomenal levels encompassing approximately two thirds of the current world population (Statista 2020), this has put immense pressure on organisations to adapt. The small device has become a significant fixture of current life. Consumers all across the globe utilise their smartphones for more than just communications, but to perform activities such as paying bills, shopping, and controlling other connected devices.

There is an evolving nature with smartphone developments and technologies as to consumer demands. Throughout the years, the digital literacy of consumers has significantly increased. With the ability to research more, connect more and ask more, consumers also want more from their smartphones. Deloitte has been surveying consumer attitudes towards smartphones

N. Wood: Ethical Considerations of AR Applications in Smartphones; A Systematic Literature Review of Consumer Perspectives    11

for the past 6 years in mainly developed markets involving 5 continents, 31 countries and 53,000 respondents (Wigginton 2018). They evaluated that more than a third of all consumers check their smartphones within 5 minutes of walking up. They confirm that mobile has become increasingly indispensable and pervasive. As such, expectations and perceptions have changed as organisations are forced to operate within this digital realm and medium.

Groß (2014) performed an empirical investigation using a modified technology acceptance model (TAM), on smartphone users in Germany to determine a consumer's intention to engage in mobile shopping. The traditional TAM factors such as ease of use and perceived usefulness were not the only factors contributing to buying intention. The results of the study found that trust and perceived enjoyment were also contributing factors in influencing a consumer's intention to mobile shop. This modified TAM approach in measuring mobile shopping acceptance provides an understanding as to how consumers perceive smartphones for particular buying activities.

Another study by Hwang and Lee (2020) performed eye tracking on consumers to evaluate their visual behaviour across a desktop screen vs a mobile screen pending their emotional state. This study was conducted to investigate interactive mobile shopping as this is currently in the early stages of development. The results revealed a greater attention to detail among mobile screens than that of desktop. This study also discovered that negative emotions yield a greater visual attention to information presented on mobile screen sizes.

As smartphones have expanded the realm of the internet as well as changing how people have access to information, the not so positive effects of smartphones has been evaluated by Kushlev and Proulx (2016). Using data from the World Values Survey, they were able to deduce that consuming greater amounts of information on mobile phones results in lower trust with people from other religions and nationalities, strangers and neighbours. They warn of the potential unforeseen costs of convenient mobile information access towards trust in one another. They also found a higher trust was had in the above groups if information was derived from any other medium such as newspaper, TV or radio.

Although consumers may develop less trust in others, Melumad and Meyer (2020) found that consumers had a higher self-disclosure on smartphones than on a personal computer. The results of their studies determined the willingness to self-disclose was influenced by two psychological effects pertaining to the device. These were the associations of comfort that many have with their smartphones, and the narrow attention tendency to focus on the task at hand due to the device smallness when generating content. As with Kushlev and Proulx, the authors of this study also suggest implications. They advise on how this level of disclosure can influence marketing efforts with emerging technologies.

## 2.4 Ethical Concerns with Digital Technologies

With ethical philosophy having begun with Socrates in the fifth century BCE, it has been a well-studied topic whereby to enable humans to seek the need for a rational analysis of their own principles and practices (MacIntyre 2003). However, modern research has studied ethics of which extend to digital technologies. As the types of digital technologies are broad, it is a dynamic construct as new advances and technologies are brought to society.

A study by Slater et al. (2020) aims to examine and outline ethical considerations in relation to the realism of augmented reality and virtual reality specific applications. They examine several key constructs such as motion sickness, intensification of an experience, information overload, negatively affecting an individual's coping mechanisms thus instigating adverse reactions, and individuals readjusting to the real-world following application use of this type of platform. They also address concerns for attention to be paid towards the presentations of violence or abusive behaviours. They raise concerns in that although an individual may be able to overlay elements visually to the real world, this may have emotional, cognitive and behavioural effects to society. An example is these applications being used to benefit racial discrimination also has the same possibility for this to cause harm. Other concerns are the potential for virtual violence and if the same constructs of ethical considerations apply in digitally enhanced worlds. An example is acts of animal cruelty in an AR application of which the user might not do in real life. Other such acts of virtual violence may have social consequences.

As we move towards *Society 5.0*, a new vision of civilisation that merges cyberspace and physical space to reframe our relationships with technology and society, there is an abundantly greater scale of ethical concerns among this new way of interacting (Deguchi et al. 2020). Although we are already familiar experiencing elements of a merged digital and physical world such as an air conditioner automatically adjusting room temperature, Society 5.0 is the integrated nature of many digital technologies that not only guide operations, but to ultimately shape our actions and behaviour. This new society relies heavily on gathering and analysing large quantities of real-world data through artificial intelligence (AI) and advanced IT systems.

Safdar, Banja, and Meltzer (2020) propose some dangers of data processing with AI and machine learning (ML) technology. They revealed limitations that selection bias in AI applications is common and is particularly noted in facial recognition. If datasets obtained fail to incorporate underrepresented groups, when applications using this technology are applies it can have potential negative effects to society. An example is through over time forming a reliance on an automated system, can diminishes the likelihood to question erroneous results. Generalisation is also a notable ethical consideration in their study.

Other ethical concerns with digital technologies were identified within Government, particularly digital government practices. Ronzhyn and Wimmer (2019) examined technologies involved in what they term Government 3.0 such as blockchain, big data and artificial intelligence as well as technologies for data storage and service delivery. These technologies within Government 3.0 are aimed at providing an evidence-based approach to decision making and policy formation. Through examination, the ethical concerns found involved inclusivity, privacy, data use, data quality, accuracy of information, transparency, accountability, information ownership, trust, alignment of values and also cost.

Furthermore, as some existing technologies become available to commercial applications, it is important to address the ethical issues that become inherent in these types of technologies. North-Samardzic (2019) identified biometric technology such as facial recognition has implications for organisations in relation to business ethics. The themes identified revolve around security and privacy of this sensitive data as well as informed consent, regulatory frameworks and guidelines, and discrimination.

With the current literature surrounding ethics in digital technologies, there lies a gap in the nature of how these ethical considerations are formulated. Often this is suggested by the researchers or developers of these technologies. There is a need to explore a consumer's perceptions of ethics in this area involving digital technologies, namely the growing arena of augmented reality.

# 3. Methodology

The methodology section of this systematic literature review will outline the research questions and discuss the chosen research methodology applied to this study. A discussion on the advantages and disadvantages of using secondary data of a systematic literature review will also be addressed. Furthermore, the data collection technique, as well as the sampling methods will be detailed. This includes search criteria and search terms used, inclusion and exclusion criteria, the ABDC journal list criteria in using secondary data and also the ethics in the data collected. The final list of articles for data analysis will be identified according to the above methods.

## 3.1 Research Questions

*Overall research question: What are the consumer perceived ethical considerations and association towards them in smartphone AR applications?*

A further detailed list of research questions identified in the proposal for this study are listed below:

(RQ1) What are the current consumer perceptions towards embracing AR applications on smartphone?

(RQ2) What are the consumer perceived benefits in using AR applications and their association?

(RQ3) What are the consumer perceived ethical considerations in using AR applications and their association?

## 3.2 Research Methodology

The methodology for this study is by systematic literature review. As defined by Dewey and Drahota (2016), a systematic literature review enables the critical appraisal of research to answer a formulated question, through identification and selection of current research. The mechanism of evaluating current literature for this study was conducted using internet research through search engines and databases publications.

A qualitative research method was also used in this study for the evaluation of literature. The nature of a qualitative approach is suitable for the research questions as the central factor within this study is a consumer's perception. As such an exploratory approach has also been adopted in conducting the research. This theory is best suited to a flexible interpretation as is the nature of perception. George (2021) advises that exploratory methods are suitable when an area of study is fairly new. As the development of AR applications is growing in nature towards consumer usage, this field of study is relatively new, therefore the above research methodologies were best suited to this study.

### 3.3 Research Design

A systematic literature review was chosen as the research design as it can add to the existing knowledge on a topic and identify through exploration any gaps in the current knowledge (Saunders and Lewis 2018). In future, primary studies can be conducted as a result of this SLR with a thorough understanding of the current awareness around this topic. The research design for this study was conducted using secondary data as is the nature of a *systematic literature review (SLR) and uses a qualitative exploratory method*. Adopting this nature through an exploratory design, enables the researcher to assess topics in a new light through seeking new insights (Saunders and Lewis 2018). The scholars Følstad and Kvale formed the guide to this SLR process from their work through a systematic literature review on customer journeys (2018), of which a similar research process was followed. The purpose of this design facilitates the overall goal of this research objective, which is to provide an understanding and ability for business related individuals to effectively implement and deploy mobile AR related applications towards consumers, with consideration to their perceptions of the ethical considerations with this technology. As with primary data collection there are positive and negatives, these will be addressed in the following section in relation to using secondary data with systematic literature reviews.

#### 3.3.1 Advantages of Secondary Data

- Ease of access and time saving.
- Lower barrier to entry via cost/some freely available literature.
- Previous literature can span across a period of time giving a longitudinal analysis.
- Mostly accessible to anyone.
- Large amounts of secondary data for various sources.
- The ability to refine searches by changing search terms.
- Analysis can begin faster as gathering data can be done relatively quickly.
- Researchers can determine trends in historical data.

### 3.3.2 Disadvantages of Secondary Data

- As online information is broad it does not mean all literature on the topic is available online and may pose limitations in finding this literature.
- Data may not be suitable to the specific area of research.
- Secondary data may rapidly become outdated.
- Analysis of data accuracy may be required.
- The reliability to the data needs to be assesses.
- The control of the data quality is not in the hands of the current researcher.
- As data may be freely available online it reduces potential 'competitive advantages' if this data is highly valuable for a particular objective.

### 3.4 Sourcing of Selected Articles

The secondary data used in this research is sourced from web information available in the form of business reports, encyclopedias, government statistics, white papers and publications in databases and journals as well as some grey literature. The databases for searching were:

- Google Scholar – due to the vast availability of literature online.
- ProQuest database – relevant literature to marketing, business and technology.
- Science Direct – a well-established database with over 4,000 academic journals.

### 3.5 Refining Articles for Critical Review

In order to refine the data to a select size for analysis in the critical review, a process of sampling techniques was conducted using search criteria, search terms, an inclusion and exclusion criteria, with consideration to the ABDC list of journals. The ethical consideration in the data selected for this review is also outlined. This process aims to identify quality, accuracy, reliability and validity of the secondary data in this study for the purpose of addressing the research questions. This form of sampling is a non-probability technique. An initial collected list of 92 articles (Appendix 1 Table 1) were added to an excel sheet where upon completion of the refining process resulted in a total of 12 articles for data analysis. The process of selection via criteria are discussed further below. The initial 92 articles were chosen according to relevancy of search criteria and search terms among the database research, however further filtered by an inclusion and exclusion criteria and the ABDC journal list criteria.

### 3.5.1 Search Criteria

Search criteria involved a web-based approach through desk work. Initial research was conducted through Google Scholar, ProQuest and Science Direct databases. An example of results appearing on Google Scholar including the terms 'consumer perceptions of ethics in augmented reality applications' yielded 32,200 article entries. However, not all search terms were displayed in article titles from the search phrase. This was done to scope the pool of available resources on different databases. Similar numbers appeared in the other databases. The identification of limiting aspects were evaluated where the search terms had to be further broken down into sub sections. Such as 'consumer perceptions of augmented reality', 'consumer perceptions of digital ethics', etc. A list of search terms is detailed below.

### 3.5.2 Search Terms

Variations with combinations of the search terms were utilised in finding the initial 92 articles deemed most relevant to the research topic. They involved terms such as:

- Augmented reality, consumer perceptions, ethics, digital ethics, mobile applications, AR applications, smartphones, ethical considerations, ethical issues, technology, perceive, online.

### 3.5.3 Inclusion and Exclusion Criteria

In order to refine the identified 92 articles, this was done through an inclusion and exclusion criteria developed by the researcher as most appropriate for the objective of this study. The criteria are outlined below in Table 1 – Inclusion Criteria and in *Table 2 – Exclusion Criteria*:

| Inclusion | Results |
| --- | --- |
| Written in English | 92 |
| Within the last 10 years, 2012 to 2022 | 64 |
| Listed on the ABDC journal list | 19 |

*Table 1 – Inclusion Criteria*

It can be noted that the majority of articles published using the search terms and inclusion criteria were between the years 2018 to 2022 being 32 articles. 2019 and 2021 both had the highest count of published articles in their respective years being 9 per year. This reveals a growing trend in papers published in recent years reflective of the trend in growing technologies.

| Exclusion | Results |
|---|---|
| Published prior to 2012 | 28 |

*Table 2 – Exclusion Criteria*

### 3.5.4 ABDC Journal List Criteria

The ABDC journal list was formulated by the Australian Business Deans Council as a collective voice for Australian university business schools. The list ranks different journals into 4 categories of quality being A*, A, B and C (Australian Business Deans Council 2019). The current list used is the 2019 version.

In refining the 92 articles from the inclusion and exclusion lists as above *(Appendix 1, Table 3 - Initial 92 Articles)* and in consideration to the journals being ranked and included in the ABDC Journal List – this was filtered to a total of 19 articles *(Appendix 1, Table 4 – Refined 19 Articles)*

### 3.5.5 Ethics in Selected Data

In assemblage of the refined and selected articles, the researcher ensured ethical contributions to the authors were made easily accessible and available. For all 92 articles direct links to the articles are included in Table 1. In referencing the 19 articles to a representative selection of available literature for data analysis, the acknowledgement of all authors has been made in Chicago referencing style as per the requirements of the university. In utilising other researchers work for the purposes of this systematic literature review, this was taken into consideration to maintain a professional level of objectivity in analysing secondary data.

### 3.6 Selected Articles

A final of 12 articles were chosen for data analysis from the pool of relevant literature. These articles represent a suitable number of the available literature to evaluate and also adhere to the selection process. This final number was decided upon by the researcher based on relevancy, subject matter and breadth of the topic in order to explore some of the current secondary and credible data available for analysis. The select 12 articles were chosen from the 19 suitable articles and also include 1 grey literature article from Harvard Business

Review. This particular article was chosen to include in the data analysis as it pertains strongly to the research question. The Harvard Business Review is an established platform and gives a wider perspective to the research topic. The full list of these 12 articles is listed in *(Appendix 1, Table 5 – Selected 12 Articles).*

# 4. Results

Within the data analysis section of this study, it summarises the select representative pool of literature with a critical review. It will commence with a description of the results obtained through the selection criteria for the identified 12 articles. This will be followed with a critical review of the articles pertaining to the research questions.

## 4.1 Description of Articles

A complete list of the selected 12 articles can be seen in *Table 6 - Article Sample List*. Of the 12 articles, 11 of there were peer reviewed and 1 consisted of a grey literature piece from Harvard Business Review. The journal articles represented the association between consumer perceptions, ethical concerns, adoption of augmented reality, as well as discussions on government and legal responsibilities for immersive technologies inclusive of augmented reality. The empirical studies involved all adult participants and were both male and female. The studies did not include children or the elderly. Most of the research studies were spanned across ages from 18 to 65, with one study particularly focused on millennials only. All studies were published in English. The primary studies involved participant numbers of greater than 100 with 1 study of over 1000 participants and another with less than 100 participants. Varying geolocations, financial and social demographics of participants were observed in these studies.

| Author | Year | Title | Objective | Methodology | Outcome |
|---|---|---|---|---|---|
| Fox, Lynn, and Rosati | 2022 | Enhancing consumer perceptions of privacy and trust: a GDPR label perspective | Investigates consumers' privacy perceptions on perceived trustworthiness and willingness to interact with an organisation. | Empirical study. Primary data, qualitative. 389 participants. | Existing methods to communicate with consumers fail to foster positive privacy perceptions and are not compliant. |
| Alimamy and Nadeem | 2022 | Is this real? Cocreation of value through authentic experiential augmented reality: the mediating effect of perceived ethics and customer engagement | Explore the impact of authentic experiences on customers' intention to cocreate value while considering the mediating influence of perceived ethics. | Empirical study. Primary data, quantitative. 266 participants. | AR generates perceptions of authentic experiences. AR increases customer perceptions of ethics and customer engagement. |
| Qin, Peak, and Prybutok | 2021 | A virtual market in your pocket: How does mobile augmented reality (MAR) influence consumer decision making? | Investigates the extent to which Mobile Augmented Reality (MAR) apps can influence user attitudes and shopping behaviour. | Empirical study. Primary data, quantitative. 162 participants | This research finds that consumers who possess positive attitudes are more willing to use a MAR app again. |

| Author | Year | Title | Purpose | Methodology | Key Findings |
|---|---|---|---|---|---|
| Suh and Prophet | 2017 | The state of immersive technology research: A literature analysis | To understand current knowledge and what we need to know about immersive technology and how users experience these technologies. | Systematic Literature Review. Secondary data of 54 articles. | The use of immersive technology can engender positive and negative consequences. |
| Fullerton, Brooksbank, and Neale | 2017 | Consumer perspectives on the ethics of an array of technology-based marketing strategies: An exploratory study | To identify consumer perceived ethics across an array of technology-based marketing initiatives. | Empirical study. Primary data, qualitative. 967 participants. | 3 sub-dimensions of the ethics related for technology-based marketing initiatives were identified: involvement, communication, and privacy. |
| Bucic, Harris, and Arli | 2012 | Ethical Consumers Among the Millennials: A Cross-National Study | Investigate the powerful, unique millennial consumer group and their engagement in ethical consumerism. | Empirical study. Primary data, quantitative. 1241 participants. | Health is the universal concern in ethical consumerism. Greater awareness does not always lead to greater purchase frequency. |
| Kamalul Ariffin, Mohan, and Goh | 2018 | Influence of consumers' perceived risk on consumers' online purchase intention | To examine the relationship between six factors of consumers' perceived risk and consumers' online purchase intentions. | Empirical study. Primary data, quantitative. 350 participants. | Results indicate 5 of the 6 factors negatively impact consumer online purchase intentions: financial risk, product risk, security risk, time risk and psychological risk. |
| Bye et al. | 2019 | The ethical and privacy implications of mixed reality | The panel explores implications of biometric data to contextually aware computing. | Conference Proceedings | Accessibility for mixed reality needs to be improved for elderly & disabled. Physical and educational access. Privacy needs a legal definition of context. |
| Royakkers et al. | 2018 | Societal and ethical issues of digitization | To discusses the social and ethical issues based on 6 dominant technologies: Internet of Things, robotics, biometrics, persuasive technology, virtual & augmented reality and digital platforms. | Literature Review | The new wave of digitization is putting pressure on public values of privacy, autonomy, security, human dignity, justice, and balance of power. |
| Franks | 2017 | The Desert of the Unreal: Inequality in Virtual and Augmented Reality | This paper discusses empathy, surveillance, virtual violence and equality factors in VR/AR applications. | Literature Review | Virtual and augmented reality technologies should be critically evaluated to assess their likely impact on inequality and consequences for legal and social policy. |
| Moraes | 2017 | Consumers' Concerns with How They Are Researched Online | To understand ethical concerns with consumers through online research. | Empirical study. Primary data, qualitative. Four qualitative focus groups, 28 participants. | Care and political responsibility should be a framework for online consumer research. |
| Javornik et al. | 2021 | Research: How AR Filters Impact People's Self-Image | Discuss potential harms of AR towards consumers mental well-being and the development of a code of ethics to AR application development. | Grey Literature | Organisations must be proactive about addressing unforeseen challenges that with augmenting a person's sense of self. |

*Table 6 - Article Sample List (Selected 12 Articles for Data Analysis).*

The majority of the studies examined ethical considerations of emerging technologies which included augmented reality. Some studies explored a more societal view and discussion on the ethical concerns from a business and industry perspective. Other literature performed primary data collection methods for a direct analysis of examination and investigation of participants perspectives and experiences. Of the 12 articles there were 7 empirical studies, 1 systematic literature review, 1 conference proceeding and 3 literature reviews. Of the empirical studies most were performed quantitatively.

# 5. Discussion

## 5.1 Consumer Perceptions Towards Using AR Applications

Some researchers indicate that the mediating influence of perceived ethics is a contributor towards cocreation when using augmented reality applications (Alimamy and Nadeem 2021). In evaluating if AR applications are embraced by consumers and the current perceptions towards using this technology, they conducted a study in which participants were asked to try the 'IKEA PLACE' AR app. Results of the online survey revealed that although AR generated perceptions of authentic experiences, it did not increase intention to cocreate value. Interestingly, they found that the authentic experiences increased the perceptions of ethics and engagement. These in turn lead to an increased intention to cocreate. Their research indicates an extension to the current understanding of technology within the cocreation process. The engagement factor of this study is further extended by Fox, Lynn, and Rosati (2022). Their study was conducted with 389 participants of which consisted of a 2 part study. Study 1 involved the perception of risk, control and privacy. Study 2 involved how these perceptions of privacy related to a consumer's trustworthiness and willingness to engage with an organisation. The findings show a support for transparent communications where explicit consent positively influences a consumer's willingness to interact and/or disclose information to organisations online.

Seen as a knowledge gap by Moraes (2017), is the understanding from a consumer's perspective on how they are researched online for use by marketers for online marketing practices. These marketing practices are diverse and broad however the exploration of online research means the 'virtual extensions of our everyday lives' (Marlowe and Allen 2022). This study was conducted through small focus groups to enable participants to shed light on their attitudes and behaviours surrounding their perspective. The participants were young professionals who were savvy technology users including social media applications on smartphones. The finding suggests online privacy is a strong performer in consumer perceptions when being researched online such as online identities. Consumer vulnerability was paradoxical such as limited digital literacy in knowing how T&C's (Terms and Conditions) actually work, while also having an implied trust in regulatory powers, yet there are limited regulations online. Consumers also revealed somewhat caring about anonymity and consent, while having a notion of self-responsibility for limited digital literacy. It should be noted, this empirical study was performed with the lowest number of participants from all the selected articles critically analysed in this literature review. With a total of 28 participants, it may not provide a conclusive set of findings. The implied trust of regulatory powers could be influenced by the implied trust consumers have of their smartphones and the miserliness nature of offloading analytical thinking to smartphones (Barr et al. 2015).

A conference proceeding on the ethical and privacy implications of mixed reality was conducted in 2019 at the SIGGRAPH conference (Bye et al. 2019). A panel of 5 members spoke over a 1.5 hour segment exploring perspectives on the implications of biometric data for virtual and augmented reality. With invitation from the audience to address matters, some of the key points in this conference were the accessibility for mixed reality needs to be improved for the elderly and disabled. This can include physical access mechanism as well as educational access for those less accustomed to the platforms on which these technologies are delivered, such as smartphones, headsets or VR glasses. Another outcome of this proceeding was the identified ambiguity on political responsibilities for the ethical and privacy concerns for AR and VR applications. This ambiguity was suggested as causation for lack of a legal definition of privacy in the context of virtual and augmented reality.

### 5.2 Consumer Benefits of AR Applications and Association with Businesses

As digital transformation puts pressure on businesses to keep abreast of emerging technology and the growth in augmented reality applications, Qin, Peak, and Prybutok (2021) investigated elements of this phenomenon. They found how mobile augmented reality apps influence consumer attitudes which in turn affect their purchasing behaviours. With 162 participants, they concluded that if the mobile augmented reality app contributed to benefit a consumer through purposeful information acquisition, enjoyment and usefulness, a positive association towards the app was found. Gratification and informativeness for the consumer are essential factors that should be considered in developing a business marketing strategy for mobile AR apps. This notion of AR apps being of benefit to a consumer was further researched in the grey literature by Javornik et al. (2021). This grey literature was included in this critical analysis of data as it can reduce publication bias, and also contribute to a comprehensive representation of available evidence (Paez 2017). It is considered more an application to theory rather than the developments of theory. This Harvard Business Review article discusses how augmented reality apps can offer highly personalised and interactive experiences, however for some purposes such as virtual try on makeup, can have a negative impact on a consumers phycological well-being (Javornik et al. 2021). The findings suggest that a baseline users current self-esteem levels influence the outcome of benefits from using the app. The researchers suggested several strategies for organisations to consider upon deployment of these technologies in their marketing strategies. They are to avoid unrealistic stands of self, be proactive in educating users of these apps on the potential harms of AR, and to develop a code of ethics with regulatory bodies and industry leaders as a developmental guide.

Many of the studies investigated the ethical consumerism in engagement with digital technologies. Not all of these studies involved the use of augmented reality. However these pieces of literature did evaluate the considerations that consumers have in their attitudes towards engaging with an organisation such as purchase intentions. Kamalul Ariffin, Mohan, and Goh (2018) sough to understand the relationship between six factors of risk. These include financial risk, product risk, security risk, time risk, social risk and psychological risk. Five of the six risk factors had a negative impact on the outcome of purchase intentions, where social risk was considered inconsequential. Security risk was found to be the greatest deterring factor to engaging with an organisation. Bucic, Harris, and Arli (2012) further evaluate the engagement with brands of millennial consumers in regard to ethical consumerism. They found through evaluating hundreds of Australian and Indonesian participants, the results varied according to country of residence. Australians who had a positive association with ethical concerns towards a brand were more likely to actionably engage, however this did not reflect the same with Indonesian participants. Most participants mentioned that increased interest in ethical issues was due to more information about the topics being available. If organisations market to an individual's sense of morals, consumers may affect change through purchasing habits (Irwin 2015).

## 5.3 Ethical Deliberations Considered by Consumers with Augmented Reality

Fullerton, Brooksbank, and Neale (2017) explored the consumer perspectives of ethics on various technology-based marketing strategies. After 967 individuals participated in their study across 20 technology-based initiatives, consumers deemed more than half of these unacceptable forms of marketing from companies. The most ethical constructs found were those considering involvement, communication and privacy. Recent statistics have also found that in emerging from the pandemic (Covid 19), post-pandemic privacy is a greater concern among Australian consumers than it was prior to the pandemic (Corbett and Shedden 2021).

In the systematic literature review by Suh and Prophet (2018), they found among 54 articles about current knowledge of immersive technologies and how users experience these, that both positive and negative consequences are engendered. They also identified that studies within this realm of virtual and augmented realities are predominantly performed on student participants. Franks (2017) goes on to explore the factors of inequality through AR and VR (virtual reality) applications in her literature review. She argues that existing 'real world' forms of inequality tend to be replicated in these immersive forms of technology. She highlights that for these technologies to truly be successful, they should be revaluated from a design perspective and a social and regulatory standpoint. With current studies finding that virtual and augmented reality assist in healthcare and patient recovery (Mubin et al. 2019), VR and AR can be positively leveraged as forces of equality for perception-expansion of individuals. A similar literature review by Royakkers et al. (2018) evaluated social and ethical issues based across 6 digital technologies including virtual and augmented reality. They found recurring themes of privacy, autonomy, security, human dignity, justice, and the balance of power to be the dominant values pressured by the increase in digitisation. They assess the strongest areas of improvement have been the privacy and security arena, however greater management of the other values are yet to be addressed in the attempt to shape society ethically.

### 5.4 Trends in Critical Analysis

In analysing the selected data, common themes were found among the articles in relation to the consumer perceptions as per the research questions, with identified key factors in using augmented reality applications and their association either positive, negative or neutral. The below figure demonstrates this association.

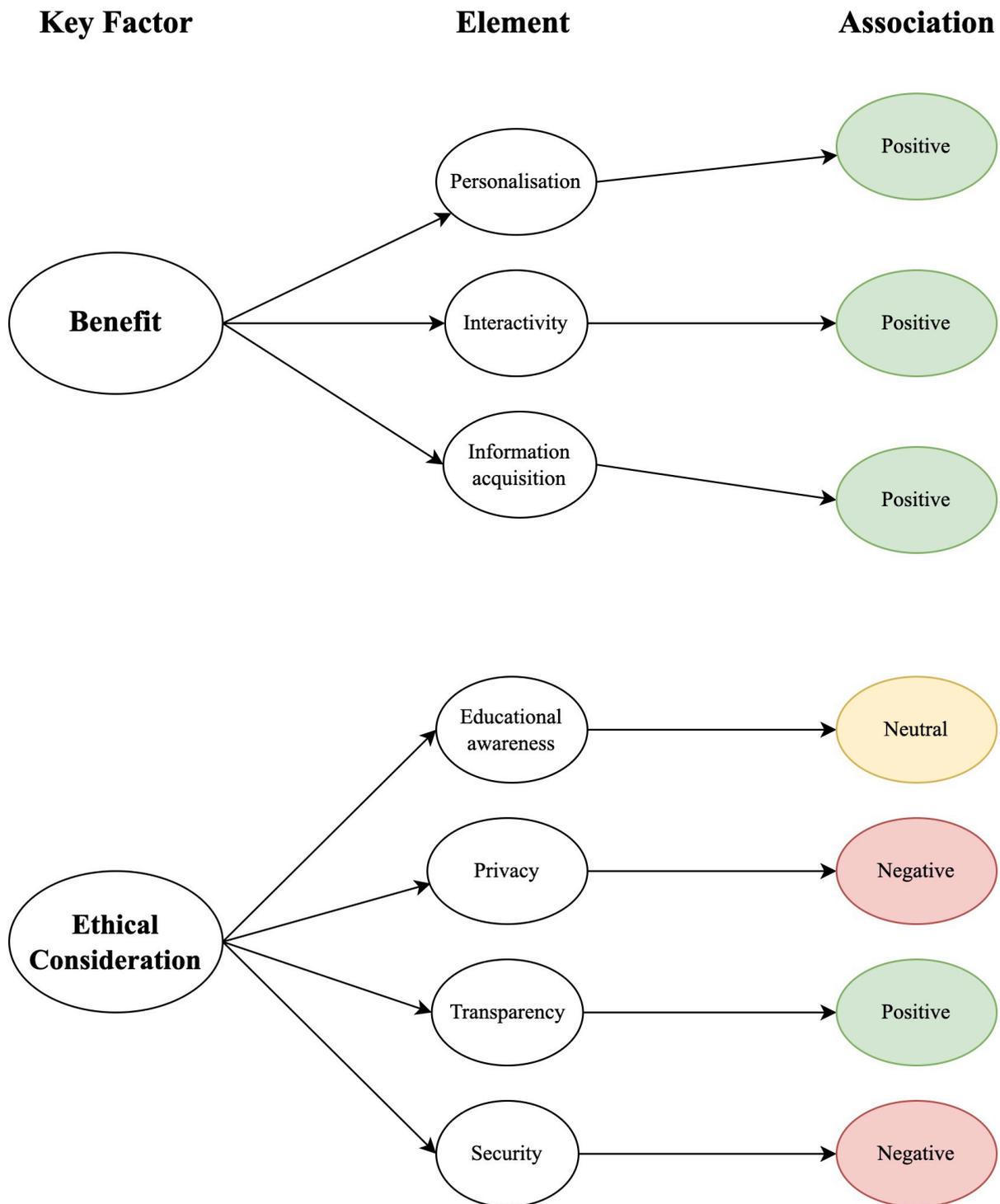

*Figure 1. Consumer perceptions and associations with identified elements*

The results suggest that consumers do consider ethics when engaging with AR technologies. Insights were extracted from the collected research with trends of ethical considerations identified upon the critical analysis of the selected articles. These ethical considerations consisted of the following 4 frequently occurring themes: educational awareness, privacy, transparency and security. Their association to these categories was determined as negative, neutral or positive. From an ethics standpoint consumers view privacy and security negatively. They view transparency in a positive light and have a minimal, often fairly muted association with current educational awareness as an ethical consideration of AR applications. It would be of value to investigate with further studies into why transparency was the only positively associated ethical consideration a consumer perceived. The reasoning behind how transparency is singularly associated as positive in respect to the other ethical considerations could provide significant value for further business practices and professionals. It can also uncover if perhaps methods of deployment or strategic campaigning have assisted in this outcome of perception.

In addition to these ethical considerations, key benefits were also identified in a similar nature with 3 frequent elements occurring and categorised into themes of personalisation, interactivity and information acquisition. The association with all 3 of these benefits was positive and further highlighted specific adoption factors to embracing this technology from a consumer perception.

Other trends among the literature found consumers to an extent do consider some of the ethical deliberations when using digital technologies, however are influenced greatly by a person's sense of self, country, cultural value system as well as other factors such as personal benefit. The specific content of using augmented reality systems on smartphones is minimally researched from a consumer's perspective. Education and information availability to the ethical considerations is an important factor in a consumer's ability to make an ethically informed decision. Benefits derived from using augmented reality applications are those seen which involve a positive user experience and personalisation. Privacy and security are prevalent concerns that consumers have of digital technologies, specifically data acquisitions and usage derived from AR applications. Concerns from researchers on the longer-term effects of augmented reality from a legal and regulatory standpoint, and the effects on society are often acknowledged. Privacy and security are recurring themes, and such represent a dominant perception of ethics among consumers regarding augmented reality.

Another trend found within the articles is a focus on millennials or those users who have experience in using digital technologies, such as smart phones or the internet. In considering this trend, the focus may not take a wholesome approach to the general populace as children were not included in any of the studies and minimally included for the elderly.

## 5.5 Implications and Study Limitations

Theoretical implications to the findings may suggest that ethical perceptions of augmented reality applications could vary upon the demographics and sociographic of users. Therefore some identified models of ethics may not apply to certain groups of individuals. With ambiguity surrounding ethical regulations on augmented reality at present, this may leave room for exploitation of trust with consumers and their smartphones (the common delivery of these technologies). Practical implications to this study could be in future, that although AR offers many benefits, if these are manipulated unethically by businesses or marketers, it may impact society in a way we are yet to understand. This could be through the ability to hyper personalise augmented reality applications to consumers, and the ability to persuade users to experience real-life scenarios, a mental health crisis could arise. The potential to drive social change could also be positive or negative. The use of augmented reality may also open consumers to unknown privacy and security vulnerabilities. These implications could impact businesses, society and individuals.

It should be acknowledged that the limitations to this study involve an ability to consider the viewpoint of all consumers to represent the populace with a greater accuracy, specifically those individuals under 18 years of age and the elder populations. Limitations can also be seen with many of the participants in the empirical studies having been experienced and/or frequent technology users. Other study limitations include the available literature specifically on smartphone augmented reality applications from a consumer's perspective. As many of these applications are delivered through a smartphone device (Bhorkar 2017), this assumption may be implied in the studies, however augmented reality applications can also be used across other devices such as tablets and glasses.

# 6. Conclusion

This systematic literature review aimed to explore the ethical concerns in AR applications in smartphones from a consumer perspective. Examination of available literature through several criteria to refine the literature to a relevant and peer reviewed pool of data included search, inclusion and exclusion lists, as well as the ABDC journal list criteria. A total of 12 pieces of literature were critically evaluated and this included one grey literature. Upon analysis, identified trends were found where consumers perceived security and privacy factors as leading ethical considerations in the context of augmented reality use. An understanding of the benefits a user experiences for engaging with AR applications was identified being personalisation, interactivity and information acquisition. This dynamic is also influenced by other demographics such as cultural values and origin of country. A lack of regulatory practices around augmented reality was identified where further study is recommended, as is including different age groups in empirical studies moving forward to expand perceptions.

The findings of this systematic literature review reveal several recommendations for further studies and areas of focus. For empirical studies it is recommended to include younger generations such as 16 - 18 years of age in the demographics of participants for analysis. This is due to a consistent limitation of participants in this age bracket within studies. A more concise nature of the devices used to study augmented reality applications is also recommended to determine if different devices have varying outcomes among consumers and their perceptions of ethics. A focus on the political and regulatory nature is also recommended for further research as there is a lack of solid understandings around their contributions to the responsible ethics of immersive technology as we move into the future.

Kamalul Ariffin, Shaizatulaqma, Thenmoli Mohan, and Yen-Nee Goh. 2018. "Influence of Consumers' Perceived Risk on Consumers' Online Purchase Intention." *Journal of Research in Interactive Marketing* 12 (3): 309–27. https://doi.org/10.1108/jrim-11-2017-0100.

Kaur, Inderjeet, E. Laxmi Lydia, Vinay Kumar Nassa, Bhanu Shrestha, Jamel Nebhen, Sharaf Malebary, and Gyanendra Prasad Joshi. 2021. "Generative Adversarial Networks with Quantum Optimization Model for Mobile Edge Computing in IoT Big Data." *Wireless Personal Communications*, June. https://doi.org/10.1007/s11277-021-08706-7.

Khosla, Rajiv, Ernesto Damiani, and William Grosky. 2003. "Why Human-Centered E-Business?" *Human-Centered E-Business*, 1–12. https://doi.org/10.1007/978-1-4615-0445-0_1.

Kushlev, Kostadin, and Jason D. E. Proulx. 2016. "The Social Costs of Ubiquitous Information: Consuming Information on Mobile Phones Is Associated with Lower Trust." Edited by Eldad Yechiam. *PLOS ONE* 11 (9): e0162130. https://doi.org/10.1371/journal.pone.0162130.

MacIntyre, Alasdair. 2003. *A Short History of Ethics*. Routledge. https://doi.org/10.4324/9780203131121.

Marlowe, Jay, and Jemma Allen. 2022. "Relationality and Online Interpersonal Research: Ethical, Methodological and Pragmatic Extensions." *Qualitative Social Work*, April, 147332502210879. https://doi.org/10.1177/14733250221087917.

Melumad, Shiri, and Robert Meyer. 2020. "Full Disclosure: How Smartphones Enhance Consumer Self-Disclosure." *Journal of Marketing* 84 (3): 28–45. https://doi.org/10.1177/0022242920912732.

Moraes, Caroline. 2017. "Consumers' Concerns with How They Are Researched Online." *Business and Professional Ethics Journal* 36 (1): 79–101. https://doi.org/10.5840/bpej2016122853.

Mubin, Omar, Fady Alnajjar, Nalini Jishtu, Belal Alsinglawi, and Abdullah Al Mahmud. 2019. "Exoskeletons with Virtual Reality, Augmented Reality, and Gamification for Stroke Patients' Rehabilitation: Systematic Review." *JMIR Rehabilitation and Assistive Technologies* 6 (2): e12010. https://doi.org/10.2196/12010.

North-Samardzic, Andrea. 2019. "Biometric Technology and Ethics: Beyond Security Applications." *Journal of Business Ethics*, March. https://doi.org/10.1007/s10551-019-04143-6.

Nwankpa, Joseph, and Yaman Roumani. 2016. "IT Capability and Digital Transformation: A Firm Performance Perspective Completed Research Paper." https://core.ac.uk/download/pdf/301370499.pdf.

Paez, Arsenio. 2017. "Grey Literature: An Important Resource in Systematic Reviews." *Journal of Evidence-Based Medicine* 10 (3). https://doi.org/10.1111/jebm.12265.

Priyono, Anjar, Abdul Moin, and Vera Nur Aini Oktaviani Putri. 2020. "Identifying Digital Transformation Paths in the Business Model of SMEs during the COVID-19 Pandemic." *Journal of Open Innovation: Technology, Market, and Complexity* 6 (4): 104. https://doi.org/10.3390/joitmc6040104.

Qin, Hong, Daniel Peak, and Victor Prybutok. 2021. "A Virtual Market in Your Pocket: How Does Mobile Augmented Reality (MAR) Influence Consumer Decision Making?" *Journal of Retailing and Consumer Services* 58 (January): 102337. https://doi.org/10.1016/j.jretconser.2020.102337.

Rauschnabel, Philipp A. 2021. "Augmented Reality Is Eating the Real-World! The Substitution of Physical Products by Holograms." *International Journal of Information Management* 57 (April): 102279. https://doi.org/10.1016/j.ijinfomgt.2020.102279.

# 8. Appendices

**Appendix 1**

*Table 3 – Initial 92 Articles*

| Title | Year | Journal or Publisher | Link |
|---|---|---|---|
| Enhancing consumer perceptions of privacy and trust: a GDPR label perspective | 2022 | Information Technology & People | https://www.emerald.com/insight/content/doi/10.1108/ITP-09-2021-0706/full/html |
| Is this real? Cocreation of value through authentic experiential augmented reality: the mediating effect of perceived ethics and customer engagement | 2022 | Information Technology & People | https://www.proquest.com/docview/2642844810/90521A0763CA4A28PQ/1?accountid=10381 |
| I want it my way! The effect of perceptions of personalization through augmented reality and online shopping on customer intentions to co-create value | 2022 | Computers in Human Behavior | https://doi.org/10.1016/j.chb.2021.107105 |
| World Intellectual Property Report 2022: The Direction of Innovation | 2022 | WIPO | https://www.wipo.int/edocs/pubdocs/en/wipo-pub-944-2022-en-world-intellectual-property-report-2022.pdf |
| To track or not to track: examining perceptions of online tracking for information behavior research | 2021 | Internet Research | https://www.emerald.com/insight/content/doi/10.1108/INTR-01-2021-0074/full/html |
| Consumer ethics: A review and research agenda | 2021 | Psychology and Marketing | https://onlinelibrary.wiley.com/doi/epdf/10.1002/mar.21580 |
| A virtual market in your pocket: How does mobile augmented reality (MAR) influence consumer decision making? | 2021 | Journal of Retailing and Consumer Services | https://doi.org/10.1016/j.jretconser.2020.102337 |
| Global AR and VR Consumer Solutions Market Growth (Status and Outlook) 2021-2026 | 2021 | Market Research Insights | https://www.mrinsights.biz/report/global-ar-and-vr-consumer-solutions-market-growth-275032.html |
| Balancing User Privacy and Innovation in Augmented and Virtual Reality | 2021 | ITIF Information Technology & Innovation Foundation | https://itif.org/publications/2021/03/04/balancing-user-privacy-and-innovation-augmented-and-virtual-reality |
| The personal digital twin, ethical considerations | 2021 | Philosophical Transactions of The Royal Society A | https://doi.org/10.1098/rsta.2020.0367 |
| CIRO: The Effects of Visually Diminished Real Objects on Human Perception in Handheld Augmented Reality | 2021 | Electronics | https://www.proquest.com/docview/2548384155/B491EAAB79D84861PQ/20?accountid=10382 |
| Consumers fear businesses are prioritizing speed over security as online fraud and identity theft grow: New research highlights rising concerns around online security risks as reliance on online services increases. | 2021 | PR Newswire | https://www.proquest.com/docview/2546527825/66FCB293ED854CABPQ/3?accountid=10382 |
| Research: How AR Filters Impact People's Self-Image | 2021 | Harvard Business Review | https://hbr.org/2021/12/research-how-ar-filters-impact-peoples-self-image |
| A new reality: Fan perceptions of augmented reality readiness in sport marketing | 2020 | Computers in Human Behavior | https://www.proquest.com/docview/2375824227/B491EAAB79D84861PQ/3?accountid=10382 |
| An Empirical Evidence Study of Consumer Perception and Socioeconomic Profiles for Digital Stores in Vietnam | 2020 | Sustainability | https://doi.org/10.3390/su12051716 |

| Title | Year | Journal | Link |
|---|---|---|---|
| Consumers' Perception of the Ethical Implications of the Utilization of Biometric Data in Targeted Digital Marketing | 2020 | I-Manager's Journal on Management | http://dx.doi.org/10.26634/jmgt.15.2.17359 |
| Realizing an Internet of Secure Things: A Survey on Issues and Enabling Technologies | 2020 | IEEE Communications Surveys and Tutorials | https://www.proquest.com/docview/2407037549/4BB74952FA134867PQ/18?accountid=10382 |
| Reflection of GDPR by the Czech Population | 2020 | Management & Marketing. Challenges for the Knowledge Society | https://doi.org/10.2478/mmcks-2020-0005 |
| Smartphone OS and User Emotion and Ethics | 2020 | Technical Gazette | https://hrcak.srce.hr/239094 |
| A Systematic Review of Consumer Perceptions of Smart Packaging Technologies for Food | 2020 | Frontiers in Sustainable Food Systems | https://doi.org/10.3389/fsufs.2020.00063 |
| Virtual reality, real reactions?: Comparing consumers' perceptions and shopping orientation across physical and virtual-reality retail stores | 2019 | Computers in Human Behavior | https://www.proquest.com/docview/2236168109/F0644C442AC54E4CPQ/1?accountid=10382 |
| Consumers' Perceptions of Native Advertisements - Exploring the Impact of Ethics and Ad Trust | 2019 | Business and Professional Ethics Journal | https://doi.org/10.5840/bpej201981584 |
| The ethical and privacy implications of mixed reality | 2019 | Association for Computing Machinery | https://doi.org/10.1145/3306212.3328138 |
| Impact Of Traditional and Digital Marketing On Consumer Perception | 2019 | Economic and Social Development: Book of Proceedings | https://www.proquest.com/docview/2317570850/B86BE8B8560042ECPQ/2?accountid=10382 |
| Healthcare professionals' competence in digitalisation: A systematic review | 2019 | Journal of Clinical Nursing | DOI:10.1111/jocn.14710 |
| Augmented Reality and Virtual Reality: The Power of AR and VR for Business | 2019 | Springer International Publishing AG | https://www.proquest.com/docview/2189584758/8B61466AB4B04F7FPQ/7?accountid=10382 |
| Privacy and the Ethics of Disability Research: Changing Perceptions of Privacy and Smartphone Use | 2019 | Second International Handbook of Internet Research | https://link.springer.com/referenceworkentry/10.1007/978-94-024-1555-1_66?noAccess=true |
| The Role of Consumers' Perceived Security, Perceived Control, Interface Design Features, and Conscientiousness in Continuous Use of Mobile Payment Services | 2019 | Sustainability | https://doi.org/10.3390/su11236843 |
| How Augmented Reality Affects People's Perceptions: Adoption of AR in Product Display Improves Consumers' Product Attitude | 2019 | Journal of physics | https://www.researchgate.net/publication/335194538_How_Augmented_Reality_Affects_People's_Perceptions_Adoption_of_AR_in_Product_Display_Improves_Consumers'_Product_Attitude |
| Influence of consumers' perceived risk on consumers' online purchase intention | 2018 | Journal of Research in Interactive Marketing | https://doi.org/10.1108/jrim-11-2017-0100 |
| Societal and ethical issues of digitization | 2018 | Ethics and Information Technology | https://doi.org/10.1007/s10676-018-9452-x |
| Exploring the dimensions of individual privacy concerns in relation to the Internet of Things use situations | 2018 | Digital Policy, Regulation and Governance | https://www.proquest.com/docview/2131849903/364968272FBE486APQ/3?accountid=10382 |
| The Needed Merge of Augmented Reality Smartphone Application with CAS and SDI Library Services | 2018 | IET Conference Proceedings | https://www.proquest.com/docview/2456901927/4A8122851ECC457BPQ/1?accountid=10382 |
| Mobile AR Performance Issues in a Cultural Heritage Environment | 2018 | International Journal of Creative Interfaces and Computer Graphics | https://www.proquest.com/docview/2296603372/4A8122851ECC457BPQ/5?accountid=10382 |

| Title | Year | Journal | URL |
|---|---|---|---|
| The state of immersive technology research: A literature analysis | 2017 | Computers in Human Behavior | https://www.proquest.com/docview/2093197609/3A75C32FAA304528PQ/4?accountid=10382 |
| Consumer perspectives on the ethics of an array of technology-based marketing strategies: An exploratory study | 2017 | Asia Pacific Journal of Marketing and Logistics | https://www.researchgate.net/publication/320261094_Consumer_perspectives_on_the_ethics_of_an_array_of_technology-based_marketing_strategies_An_exploratory_study |
| How augmented reality apps are accepted by consumers: A comparative analysis using scales and opinions | 2017 | Technological Forecasting and Social Change | https://www.sciencedirect.com/science/article/abs/pii/S0040162516304528 |
| Electronic cigarette retailers use Pokémon Go to market products | 2017 | Tobacco Control | https://doi.org/10.1136/tobaccocontrol-2016-053369 |
| The Desert of the Unreal: Inequality in Virtual and Augmented Reality | 2017 | SSRN | https://ssrn.com/abstract=3014529 |
| Consumers' Concerns with How They Are Researched Online | 2017 | Business and Professional Ethics Journal | https://doi.org/10.5840/bpej2016122853 |
| Why enterprises are not ready for the future | 2017 | Computerworld Hong Kong | https://www.proquest.com/docview/1933300160/936345711D4D4E62PQ/5?accountid=10382 |
| Digitization as an ethical challenge | 2017 | AI & Society | https://link.springer.com/article/10.1007/s00146-016-0686-z |
| Technologies of Nonviolence: Ethical Participatory Visual Research with Girls | 2017 | Girlhood Studies: An Interdisciplinary Journal | https://doi.org/10.3167/ghs.2017.100205 |
| Information Privacy for Technology Users with Intellectual and Developmental Disabilities: Why Does It Matter? | 2017 | Ethics & Behavior | https://www.researchgate.net/publication/320660441_Information_Privacy_for_Technology_Users_With_Intellectual_and_Developmental_Disabilities_Why_Does_It_Matter |
| A Survey on Applications of Augmented Reality | 2016 | Advances in Computer Science: an International Journal | http://www.acsij.org/acsij/article/view/400 |
| Online privacy and security concerns of consumers | 2016 | Information Management & Computer Security | https://www.proquest.com/docview/1830450039/7650935825B49CCPQ/2?accountid=10382 |
| Web tracking awareness and online privacy | 2016 | ProQuest Dissertations Publishing | https://www.proquest.com/docview/1836098615/C7B4FF5835D24574PQ/3?accountid=10382 |
| There's something in your eye: ethical implications of augmented visual field devices | 2016 | Journal of Information, Communication and Ethics in Society | https://www.emerald.com/insight/content/doi/10.1108/JICES-10-2015-0035/full/html |
| Click here to consent forever: Expiry dates for informed consent | 2016 | SAGE Journals | https://journals.sagepub.com/doi/10.1177/2053951715624935 |
| Playful Biometrics: Controversial Technology through the Lens of Play | 2016 | The Sociological Quarterly | https://www.jstor.org/stable/23027564?seq=1 |
| Consumer trust established by E-commerce privacy indicators | 2015 | ProQuest Dissertations Publishing | https://www.proquest.com/docview/1779999468/F1C638F8AB644DF7PQ/4?accountid=10382 |
| Augmented Reality Law, Privacy, and Ethics: Law, Society, and Emerging AR Technologies | 2015 | ScienceDirect | https://www.proquest.com/docview/1688713504/A9CEC3E3D717417DPQ/1?accountid=10382 |

| Title | Year | Journal | URL |
|---|---|---|---|
| Multidimensionality: redefining the digital divide in the smartphone era | 2015 | info | https://www.emerald.com/insight/content/doi/10.1108/info-09-2014-0037/full/html |
| Visualizing Big Data with augmented and virtual reality: challenges and research agenda | 2015 | Journal of Big Data | https://www.proquest.com/docview/1987954712/B71770EE32A441E4PQ/8?accountid=10382 |
| Qualitative Research Method: Grounded Theory | 2014 | International Journal of Business and Management | https://www.ccsenet.org/journal/index.php/ijbm/article/view/39643 |
| Formation of augmented-reality interactive technology's persuasive effects from the perspective of experiential value | 2014 | Internet Research | https://www.proquest.com/docview/1520332832/4FA55769BE51439CPQ/2?accountid=10382 |
| Ethical perspectives on e-commerce: an empirical investigation | 2014 | Internet Research | https://www.emerald.com/insight/content/doi/10.1108/IntR-07-2013-0162/full/html |
| Challenges and Opportunities of Lifelog Technologies: A Literature Review and Critical Analysis | 2014 | Science and Engineering Ethics | https://link.springer.com/article/10.1007/s11948-013-9456-1 |
| E-Marketing and Digital Marketing Developments | 2013 | Journal of Marketing Vistas | https://www.proquest.com/docview/1513232087/B86BE8B8560042ECPQ/5?accountid=10382 |
| The Importance of Digital Marketing. An Exploratory Study to Find the Perception and Effectiveness of Digital Marketing Amongst the Marketing Professionals in Pakistan. | 2013 | Journal of Information Systems & Operations Management | https://www.proquest.com/docview/1477205997/A803CFF24E9C4CC5PQ/1?accountid=10382 |
| Current Status, Opportunities and Challenges of Augmented Reality in Education | 2013 | Computers & Education | https://www.proquest.com/docview/1413415101/83B097C6359405FPQ/2?accountid=10382 |
| Ethical Consumers Among the Millennials: A Cross-National Study | 2012 | Journal of Business Ethics | https://doi.org/10.1007/s10551-011-1151-z |
| Privacy Problems in the Online World | 2012 | IEEE Internet Computing | https://www.proquest.com/docview/1019649640/66FCB293ED854CABPQ/5?accountid=10382 |
| Ethical Considerations in Augmented Reality Applications | 2012 | Proceedings of the International Conference on e-Learning, e-Business, Enterprise Information Systems, and e-Government (EEE) | https://www.proquest.com/docview/1349774701/90521A0763CA4A28PQ/7?accountid=10382 |
| Augmented Reality for Smartphones | 2011 | University of Bath | https://researchportal.bath.ac.uk/en/publications/augmented-reality-for-smartphones |
| Augmented reality technologies, systems and applications | 2011 | Multimedia Tools and Applications | https://www.proquest.com/docview/907934325/A75784DAA6C34B6BPQ/1?accountid=10382 |
| Consumers' perceptions of online ethics and its effects on satisfaction and loyalty | 2011 | Journal of Research in Interactive Marketing | https://www.emerald.com/insight/content/doi/10.1108/17505931111121534/full/html |

| Title | Year | Source | URL |
|---|---|---|---|
| Augmented Reality: A Sustainable Marketing Tool? | 2010 | Global Business and Management Research | https://www.proquest.com/docview/920098032/E0FC4D7C977F4EBEPQ/2?accountid=10382 |
| Ethical Behavior in the Information Age | 2009 | Knowledge Quest | https://www.proquest.com/docview/194732136/B212BDCA2FA24273PQ/4?accountid=10382 |
| Online privacy and security of Internet digital certificates: A study of the awareness, perceptions, and understanding of Internet users | 2008 | ProQuest Dissertations Publishing | https://www.proquest.com/docview/230718034/7F29823C1A6C428BPQ/2?accountid=10382 |
| Mobile Persuasion: 20 Perspectives on the Future of Behavior Change | 2007 | Stanford Captology Media | https://www.amazon.com/Mobile-Persuasion-Perspectives-Future-Behavior/dp/0979502527 |
| Augmented Reality Simulations on Handheld Computers | 2007 | Journal of the Learning Sciences | https://www.proquest.com/docview/62040005/A75784DAA6C34B6BPQ/3?accountid=10382 |
| eHealth information quality and ethics issues: an exploratory study of consumer perceptions | 2007 | International Journal of Pharmaceutical and Healthcare Marketing | https://www.emerald.com/insight/content/doi/10.1108/17506120710740261/full/html |
| Computer ethics and consumer ethics: the impact of the internet on consumers' ethical decision-making process | 2007 | Journal of Consumer Behaviour | https://doi.org/10.1002/cb.223 |
| The ethics of biometrics: the risk of social exclusion from the widespread use of electronic identification | 2007 | Science and Engineering Ethics | https://www.proquest.com/docview/1035871924 |
| Convergence and trust in eCommerce | 2006 | BT Technology Journal | https://www.proquest.com/docview/215202771/7650935825B49CCPQ/3?accountid=10382 |
| Ethics of Computer Use: A Survey of Student Attitudes | 2006 | Academy of Information and Management Sciences Journal | https://www.proquest.com/docview/214625472/B6399DC861474ECAPQ/2?accountid=10382 |
| How Social Cause Marketing Affects Consumer Perceptions | 2006 | MIT Sloan Management Review | https://centers.fuqua.duke.edu/case/wp-content/uploads/sites/7/2015/02/Article_Bloom_HowSocialCauseMarketingAffects_2006.pdf |
| Consumer Perceptions of Privacy and Security Risks for Online Shopping | 2005 | The Journal of Consumer Affairs | https://onlinelibrary.wiley.com/doi/abs/10.1111/j.1745-6606.2001.tb00101.x |
| Social, Ethical and Policy Implications of Information Technology | 2004 | Journal of Documentation | https://www.proquest.com/docview/217977770/DDF82222EDF643EEPQ/2?accountid=10382 |
| Shopping for a better world? An interpretive study of the potential for ethical consumption within the older market | 2004 | Journal of Consumer Marketing | https://www.emerald.com/insight/content/doi/10.1108/07363760410558672/full/html |

| Title | Year | Journal | Link |
|---|---|---|---|
| In ecommerce, customer trust is no longer an option: It is the requirement for success | 2002 | Quality Congress. ASQ's ... Annual Quality Congress Proceedings | https://www.proquest.com/docview/214388291/BB9ACA880626449EPQ/3?accountid=10382 |
| E-commerce, ethical commerce? | 2002 | Journal of business ethics | https://www.proquest.com/docview/39046117/7650935825B49CCPQ/5?accountid=10382 |
| Computer ethics: the influence of guidelines and universal moral beliefs | 2002 | Information Technology and People | https://www.proquest.com/docview/57588946/77F16135E5644056PQ/4?accountid=10382 |
| The Ethics of Cyberspace | 2002 | The Southern Communication Journal | https://www.proquest.com/docview/226937941/6EB3AD4FDFC940CFPQ/1?accountid=10382 |
| Computer ethics: Its birth and its future | 2001 | Ethics and Information Technology | https://www.proquest.com/docview/222210086/BAF3FCFA872044E5PQ/1?accountid=10382 |
| An investigation of Internet users' level of awareness, understanding, and overall perceptions of Internet cookies in regard to online privacy and security | 2000 | ProQuest Dissertations Publishing | https://www.proquest.com/docview/304655609/88A6AF1602A44AECPQ/3?accountid=10382 |
| Our moral condition in cyberspace | 2000 | Ethics and Information Technology | https://www.proquest.com/docview/222235854/77F16135E5644056PQ/1?accountid=10382 |
| Judgements about computer ethics: Do individual, co-worker, and company judgements differ? Do company codes make a difference? | 2000 | Journal of Business Ethics | https://www.proquest.com/docview/198028545/12B0CD01DA1F4BEAPQ/5?accountid=10382 |
| A cross-cultural study of consumer perceptions about marketing ethics | 1999 | Journal of Consumer Marketing | https://doi.org/10.1108/07363769910271496 |
| Cracking the hacker myth Computer kids learn the value of ethics in a wired world: [FINAL Edition] | 1998 | USA TODAY | https://www.proquest.com/docview/408754059/AE15184654D6475BPQ/1?accountid=10382 |
| Ethics online | 1997 | Communications of the ACM | http://doi.acm.org/10.1145/242857.242875 |

*Table 4 – Refined 19 Articles*

| Author | Year | Title | Journal | ABDC Rating | Key Findings/Purpose | Measurement |
|---|---|---|---|---|---|---|
| Fox, Lynn, and Rosati | 2022 | Enhancing consumer perceptions of privacy and trust: a GDPR label perspective | Information Technology & People | A | Findings support the potential of GDPR privacy labels for positively influencing perceptions of risk, control, privacy and trustworthiness and enhancing consumers' willingness to transact and disclose data to online organizations. | Empirical study. Study 1 examines the effects of each label on perceptions of risk, control and privacy. Study 2 investigates the influence of consumers' privacy perceptions on perceived trustworthiness and willingness to interact with the organization. |

| Author | Year | Title | Journal | Rank | Key Finding | Methodology |
|---|---|---|---|---|---|---|
| Alimamy and Nadeem | 2022 | Is this real? Cocreation of value through authentic experiential augmented reality: the mediating effect of perceived ethics and customer engagement | Information Technology & People | A | The authentic experiences generated through AR increases customer perceptions of ethics and customer engagement, both of which lead to an increased intention to cocreate value. | Empirical study. Online survey - responses were used in a structural equation model. |
| Alimamy and Gnoth | 2022 | I want it my way! The effect of perceptions of personalization through augmented reality and online shopping on customer intentions to co-create value | Computers in Human Behavior | A | Perceived personalization leads to heightened perceptions of risk and trust in both AR, and online environments. | Empirical study. Online survey - data collected a total of 837 responses. |
| Makhortykh et al. | 2021 | To track or not to track: examining perceptions of online tracking for information behavior research | Internet Research | A | Examine perceptions of online tracking for information behaviour research. | Empirical study. Focus groups. |
| Hassan, Rahman, and Paul | 2021 | Consumer ethics: A review and research agenda | Psychology and Marketing | A | A review on the moral ethics of consumers. | Literature Review |
| Qin, Peak, and Prybutok | 2021 | A virtual market in your pocket: How does mobile augmented reality (MAR) influence consumer decision making? | Journal of Retailing and Consumer Services | A | This research finds that consumers who possess positive attitudes are more willing to use a MAR app again. | Empirical study of 162 participants, to investigates the extent to which Mobile Augmented Reality (MAR) apps can influence user attitudes and shopping behaviour. |
| Goebert and Greenhalgh | 2020 | A new reality: Fan perceptions of augmented reality readiness in sport marketing | Computers in Human Behavior | A | Visual appeal should be emphasized when developing AR activations in sport settings among other findings. | Empirical study. 1373 participants in an online survey. Measuring the fan experience when adopting augmented reality (AR). |
| Pizzi et al | 2019 | Virtual reality, real reactions?: Comparing consumers' perceptions and shopping orientation across physical and virtual-reality retail stores | Computers in Human Behavior | A | VR is a reliable tool for estimating consumers' in-store perceptions and behaviours among other findings. | Quasi-experimental design was implemented to test the research hypotheses, comprised two conditions of grocery store shoppers: physical and VR-based. |
| Suh and Prophet | 2017 | The state of immersive technology research: A literature analysis | Computers in Human Behavior | A | The use of immersive technology can engender positive and negative consequences. | Systematic Literature Review of 54 articles. |
| Fullerton, Brooksbank, and Neale | 2017 | Consumer perspectives on the ethics of an array of technology-based marketing strategies: An exploratory study | Asia Pacific Journal of Marketing and Logistics | A | Marketers need to complete their due diligence so as to determine which potential technology-based marketing strategies will likely be viewed as acceptable or not within their target markets. | Empirical study. Survey of 967 adult participants. |

| Author | Year | Title | Journal | Rank | Purpose | Method |
|---|---|---|---|---|---|---|
| Rese et al. | 2017 | How augmented reality apps are accepted by consumers: A comparative analysis using scales and opinions | Technological Forecasting and Social Change | A | The adoption of Augmented Reality (AR) technologies in retailing is investigated. | Four experiments with a total of 978 participants. |
| Bucic, Harris, and Arli | 2012 | Ethical Consumers Among the Millennials: A Cross-National Study | Journal of Business Ethics | A | Investigates the powerful, unique Millennial consumer group and their engagement in ethical consumerism. | Empirical study. Australia and Indonesia participants, 832 and 409 samples obtained respectively. |
| Kamalul Ariffin, Mohan, and Goh | 2018 | Influence of consumers' perceived risk on consumers' online purchase intention | Journal of Research in Interactive Marketing | B | To examine the relationship between six factors of consumers' perceived risk and consumers' online purchase intentions | Empirical study. 350 respondents participated on an online survey. |
| Kirkpatrick et al. | 2017 | Electronic cigarette retailers use Pokémon Go to market products | Tobacco Control | B | Explore the augmented reality videogame Pokémon Go on smartphones, and how some companies are promoting products such as electronic cigarettes via the game. | Literature Review |
| Fischbach and Zarzosa | 2019 | Consumers' Perceptions of Native Advertisements - Exploring the Impact of Ethics and Ad Trust | Business and Professional Ethics Journal | C | To address ethical concerns and deception in native advertisements. Results uncover that consumer ethical efficacy affects their intention to share native ads through eWOM. | Empirical study. |
| Bye et al. | 2019 | The ethical and privacy implications of mixed reality | Association for Computing Machinery | C | The panel explores the many implications of biometric data to contextually aware computing. | Conference Proceedings |
| Royakkers et al | 2018 | Societal and ethical issues of digitization | Ethics and Information Technology | C | This paper discusses the social and ethical issues based on 6 dominant technologies: Internet of Things, robotics, biometrics, persuasive technology, virtual & augmented reality and digital platforms. | Literature Review |
| Franks | 2017 | The Desert of the Unreal: Inequality in Virtual and Augmented Reality | SSRN | C | This paper discusses empathy, surveillance, virtual violence and equality factors in VR/AR applications. | Literature Review |
| Moraes | 2017 | Consumers' Concerns with How They Are Researched Online | Business and Professional Ethics Journal | C | This study aims to understand ethical concerns with consumers through online research, such as e-privacy, anonymity, consent, consumer vulnerability, co-responsibility and willingness to be researched. | Empirical study. Four qualitative focus groups. |

*Table 5 – Selected 12 Articles*

| Author | Year | Title | Journal | ABDC Rating | Objective | Methodology | Outcome |
|---|---|---|---|---|---|---|---|
| Fox, Lynn, and Rosati | 2022 | Enhancing consumer perceptions of privacy and trust: a GDPR label perspective | Information Technology & People | A | Investigates consumers' privacy perceptions on perceived trustworthiness and willingness to interact with an organisation. | Empirical study. Primary data, qualitative. 389 participants. | Existing methods to communicate with consumers fail to foster positive privacy perceptions and are not compliant. |
| Alimamy and Nadeem | 2022 | Is this real? Cocreation of value through authentic experiential augmented reality: the mediating effect of perceived ethics and customer engagement | Information Technology & People | A | Explore the impact of authentic experiences on customers' intention to cocreate value while considering the mediating influence of perceived ethics. | Empirical study. Primary data, quantitative. 266 participants. | AR generates perceptions of authentic experiences. AR increases customer perceptions of ethics and customer engagement. |
| Qin, Peak, and Prybutok | 2021 | A virtual market in your pocket: How does mobile augmented reality (MAR) influence consumer decision making? | Journal of Retailing and Consumer Services | A | Investigates the extent to which Mobile Augmented Reality (MAR) apps can influence user attitudes and shopping behaviour. | Empirical study. Primary data, quantitative. 162 participants | This research finds that consumers who possess positive attitudes are more willing to use a MAR app again. |
| Suh and Prophet | 2017 | The state of immersive technology research: A literature analysis | Computers in Human Behavior | A | To understand current knowledge and what we need to know about immersive technology and how users experience these technologies. | Systematic Literature Review. Secondary data of 54 articles. | The use of immersive technology can engender positive and negative consequences. |
| Fullerton, Brooksbank, and Neale | 2017 | Consumer perspectives on the ethics of an array of technology-based marketing strategies: An exploratory study | Asia Pacific Journal of Marketing and Logistics | A | To identify consumer perceived ethics across an array of technology-based marketing initiatives. | Empirical study. Primary data, qualitative. 967 participants. | 3 sub-dimensions of the ethics related for technology-based marketing initiatives were identified: involvement, communication, and privacy. |
| Bucic, Harris, and Arli | 2012 | Ethical Consumers Among the Millennials: A Cross-National Study | Journal of Business Ethics | A | Investigate the powerful, unique millennial consumer group and their engagement in ethical consumerism. | Empirical study. Primary data, quantitative. 1241 participants. | Health is the universal concern in ethical consumerism. Greater awareness does not always lead to greater purchase frequency. |

| Author | Year | Title | Journal | Rank | Purpose | Method | Findings |
|---|---|---|---|---|---|---|---|
| Kamalul Ariffin, Mohan, and Goh | 2018 | Influence of consumers' perceived risk on consumers' online purchase intention | Journal of Research in Interactive Marketing | B | To examine the relationship between six factors of consumers' perceived risk and consumers' online purchase intentions. | Empirical study. Primary data, quantitative. 350 participants. | Results indicate 5 of the 6 factors negatively impact consumer online purchase intentions: financial risk, product risk, security risk, time risk and psychological risk. |
| Bye et al. | 2019 | The ethical and privacy implications of mixed reality | Association for Computing Machinery | C | The panel explores implications of biometric data to contextually aware computing. | Conference Proceedings | Accessibility for mixed reality needs to be improved for elderly & disabled. Physical and educational access. Privacy needs a legal definition of context. |
| Royakkers et al. | 2018 | Societal and ethical issues of digitization | Ethics and Information Technology | C | To discusses the social and ethical issues based on 6 dominant technologies: Internet of Things, robotics, biometrics, persuasive technology, virtual & augmented reality and digital platforms. | Literature Review | The new wave of digitization is putting pressure on public values of privacy, autonomy, security, human dignity, justice, and balance of power. |
| Franks | 2017 | The Desert of the Unreal: Inequality in Virtual and Augmented Reality | SSRN | C | This paper discusses empathy, surveillance, virtual violence and equality factors in VR/AR applications. | Literature Review | Virtual and augmented reality technologies should be critically evaluated to assess their likely impact on inequality and consequences for legal and social policy. |
| Moraes | 2017 | Consumers' Concerns with How They Are Researched Online | Business and Professional Ethics Journal | C | To understand ethical concerns with consumers through online research. | Empirical study. Primary data, qualitative. Four qualitative focus groups, 28 participants. | Care and political responsibility should be a framework for online consumer research. |
| Javornik et al. | 2021 | Research: How AR Filters Impact People's Self-Image | Harvard Business Review | NA | Discuss potential harms of AR towards consumers mental well-being and the development of a code of ethics to AR application development. | Grey Literature | Organisations must be proactive about addressing unforeseen challenges that with augmenting a person's sense of self. |